\documentclass[sigconf, nonacm]{acmart}
\newcommand\vldbdoi{XX.XX/XXX.XX}
\newcommand\vldbpages{XXX-XXX}
\newcommand\vldbvolume{14}
\newcommand\vldbissue{1}
\newcommand\vldbyear{2020}
\newcommand\vldbauthors{\authors}
\newcommand\vldbtitle{\shorttitle} 
\newcommand\vldbavailabilityurl{URL_TO_YOUR_ARTIFACTS}
\newcommand\vldbpagestyle{plain} 

\newcommand{\picfolder}{./}

\begin{document}
\title{Beyond Static Question Banks: Dynamic Knowledge Expansion via LLM-Automated Graph Construction and Adaptive Generation}

%%
%% The "author" command and its associated commands are used to define the authors and their affiliations.
\author{Yingquan Wang}
\affiliation{%
   \institution{Dalian University of Technology}
%   \state{Ireland}
}
\email{yingquan_w95@mail.dlut.edu.cn}

\author{Tianyu Wei}
% \orcid{0000-0002-1825-0097}
\affiliation{%
   \institution{Chengdu Technological University}
%   \country{Iceland}
}
 \email{dontcall3@qq.com}

\author{Qinsi Li}
% \orcid{0000-0001-5109-3700}
\affiliation{%
   \institution{Shenzhen Research Institute of Big Data}
 }
 \email{liqinsi@sribd.cn}

\author{Li Zeng}
% \orcid{0000-0001-5109-3700}
\affiliation{%
	\institution{Shenzhen Research Institute of Big Data}
}
\email{zengli@sribd.cn}

\begin{abstract}

Personalized education systems increasingly rely on structured knowledge representations to support adaptive learning and question generation.
However, existing approaches face two fundamental limitations.
First, constructing and maintaining knowledge graphs for educational content largely depends on manual curation, resulting in high cost and poor scalability.
Second, most personalized education systems lack effective support for state-aware and systematic reasoning over learners’ knowledge, and therefore rely on static question banks with limited adaptability.
To address these challenges, this paper proposes a Generative GraphRAG framework for automated knowledge modeling and personalized exercise generation.
It consists of two core modules.
The first module, Automated Hierarchical Knowledge Graph Constructor (Auto-HKG), leverages LLMs to automatically construct hierarchical knowledge graphs that capture structured concepts and their semantic relations from educational resources.
The second module, Cognitive GraphRAG (CG-RAG), performs graph-based reasoning over a learner mastery graph and combines it with retrieval-augmented generation to produce personalized exercises that adapt to individual learning states.
The proposed framework has been deployed in real-world educational scenarios, where it receives favorable user feedback, suggesting its potential to support practical personalized education systems.

\end{abstract}

\maketitle

%% do not modify the following VLDB block %%
%% VLDB block start %%%
\pagestyle{\vldbpagestyle}
\begingroup\small\noindent\raggedright\textbf{PVLDB Reference Format:}\\
\vldbauthors. \vldbtitle. PVLDB, \vldbvolume(\vldbissue): \vldbpages, \vldbyear.\\
\href{https://doi.org/\vldbdoi}{doi:\vldbdoi}
\endgroup
\begingroup
\renewcommand\thefootnote{}\footnote{\noindent
This work is licensed under the Creative Commons BY-NC-ND 4.0 International License. Visit \url{https://creativecommons.org/licenses/by-nc-nd/4.0/} to view a copy of this license. For any use beyond those covered by this license, obtain permission by emailing \href{mailto:info@vldb.org}{info@vldb.org}. Copyright is held by the owner/author(s). Publication rights licensed to the VLDB Endowment. \\
\raggedright Proceedings of the VLDB Endowment, Vol. \vldbvolume, No. \vldbissue\ %
ISSN 2150-8097. \\
\href{https://doi.org/\vldbdoi}{doi:\vldbdoi} \\
}\addtocounter{footnote}{-1}\endgroup
%% VLDB block end %%%

%% do not modify the following VLDB block %%
%% VLDB block start %%%
\ifdefempty{\vldbavailabilityurl}{}{
\vspace{.3cm}
\begingroup\small\noindent\raggedright\textbf{PVLDB Artifact Availability:}\\
The source code, data, and/or other artifacts have been made available at \url{\vldbavailabilityurl}.
\endgroup
}
% VLDB block end %%%

\section{Introduction}

Personalized and adaptive learning systems have become an important direction in educational data management, aiming to tailor learning content and trajectories to individual learners’ cognitive states~\cite{corbett1994knowledge, piech2015deep}.
A key enabler of such systems is the availability of structured knowledge representations that capture educational concepts, their dependencies, and the relationships between learning materials and knowledge components~\cite{dang2021constructing}.
In practice, however, constructing and maintaining such representations remains a major bottleneck.

\begin{figure}[htbp]
    \centering
    \includegraphics[width=1.0\linewidth]{\picfolder 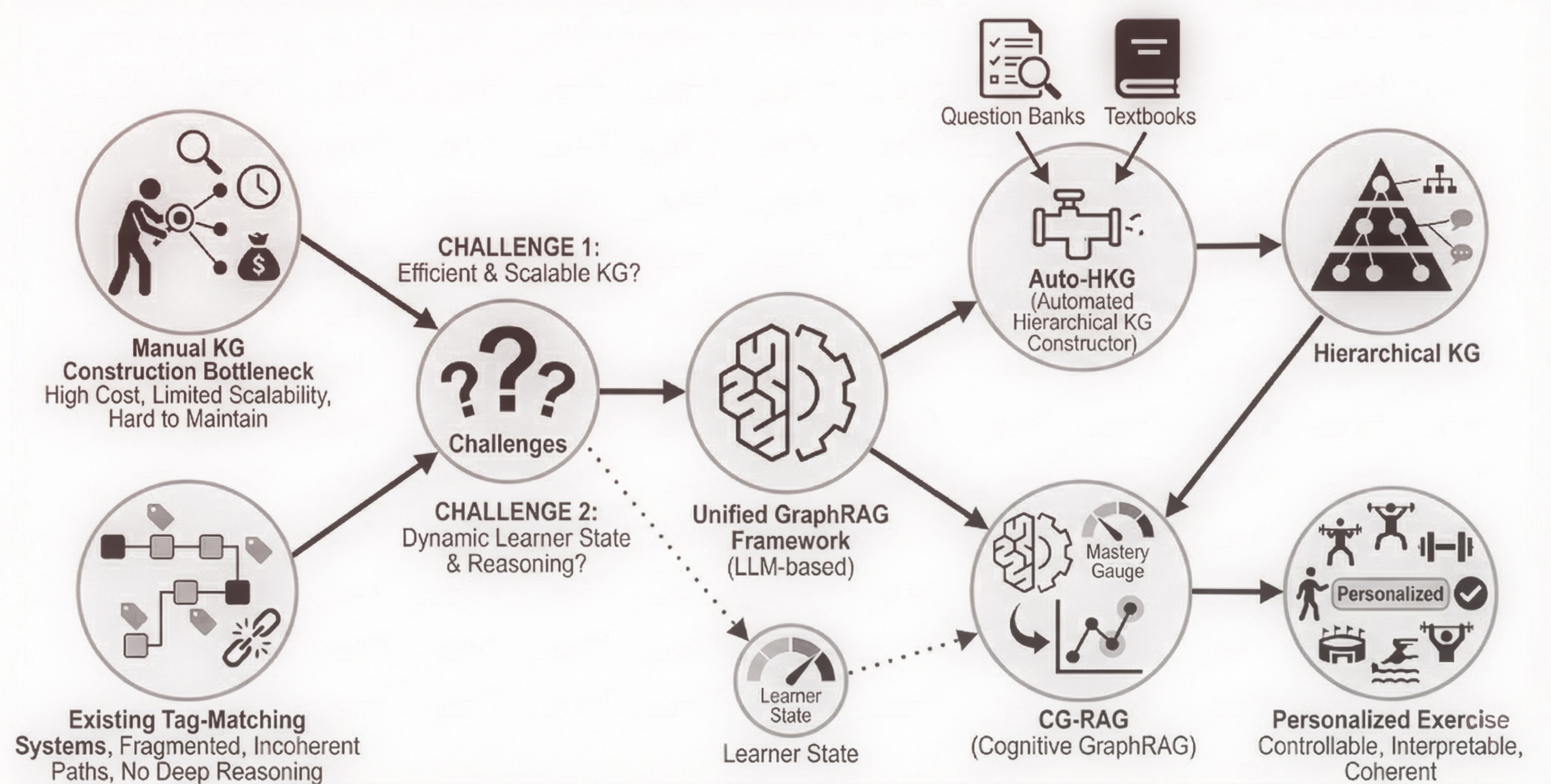}
    \caption{Overview of the proposed Generative GraphRAG framework. 
    The framework addresses two key challenges in personalized education systems: the high cost of manual knowledge graph construction, and the lack of learner-state-aware reasoning in tag-matching-based systems. 
    Auto-HKG automatically constructs a hierarchical knowledge graph from question banks and textbooks, while CG-RAG integrates learner mastery states with graph-based reasoning and retrieval-augmented generation to produce personalized exercises.
    }
    \label{fig:intro1}
\end{figure}

Most existing educational knowledge graphs~\cite{chen2018knowedu} rely heavily on expert-driven annotation, including the definition of knowledge taxonomies, prerequisite relations, and mappings between exercises and knowledge points.
This manual process incurs substantial cost, exhibits limited scalability, and is difficult to generalize across subjects, curricula, or textbook versions.
As educational content evolves through textbook revisions, changes in question formats, and the continuous expansion of question banks, the burden of manual maintenance accumulates, making it difficult for real-world systems to sustain high-quality knowledge graphs over time.
At the same time, personalized adaptive learning requires not only structured domain knowledge but also accurate modeling of learners’ dynamic cognitive states.
Many deployed systems still adopt tag-matching or rule-based recommendation strategies, where exercises are selected based on surface-level labels rather than systematic reasoning.
Such approaches often fail to capture logical dependencies among knowledge components, resulting in fragmented learning experiences and incoherent learning paths that do not adequately reflect a learner’s evolving mastery.

These limitations give rise to two fundamental challenges.
First, how can high-quality educational knowledge graphs be constructed efficiently and maintained at scale without extensive manual annotation?
Second, how can learners’ dynamic cognitive states be deeply integrated with static knowledge structures, enabling the system to reason about coherent learning trajectories and to generate or recommend personalized sets of exercises accordingly?

To address these challenges, we propose an LLM-based adaptive exercise generation system that combines automated knowledge graph construction with state-aware graph reasoning and retrieval-augmented generation.
The system first employs an Automated Hierarchical Knowledge Graph Constructor (Auto-HKG) to automatically build hierarchical knowledge graphs from existing question banks and textbook corpora.
Auto-HKG leverages structured prompting, schema constraints, and a pipeline design to capture both conceptual hierarchies and semantic relations among knowledge components.
On top of the constructed graph, we introduce Cognitive GraphRAG (CG-RAG), which maps learners’ mastery levels onto the knowledge graph and performs graph-based reasoning to guide retrieval and generation. By integrating learner states with graph-aware retrieval-augmented generation, CG-RAG produces controllable, interpretable, and coherent personalized exercise sets.

Our key contributions are as follows:
\begin{itemize}
  \item We present an adaptive exercise generation system that integrates automated knowledge graph construction, learner-state-aware graph reasoning, and LLM-based exercise generation and selection into a unified framework.
        % 是否需要强调我们首次？
  \item We propose Auto-HKG for automated hierarchical knowledge graph construction and CG-RAG for mastery-driven graph reasoning and retrieval-augmented exercise generation with dynamic updates.
  \item We validate the effectiveness of the proposed system through deployment in real-world educational scenarios, demonstrating improvements in overall user learning experience.
\end{itemize}

\section{Core Structural Elements}

\begin{figure*}[htbp]
    \centering
    \includegraphics[width=1.0\linewidth]{\picfolder 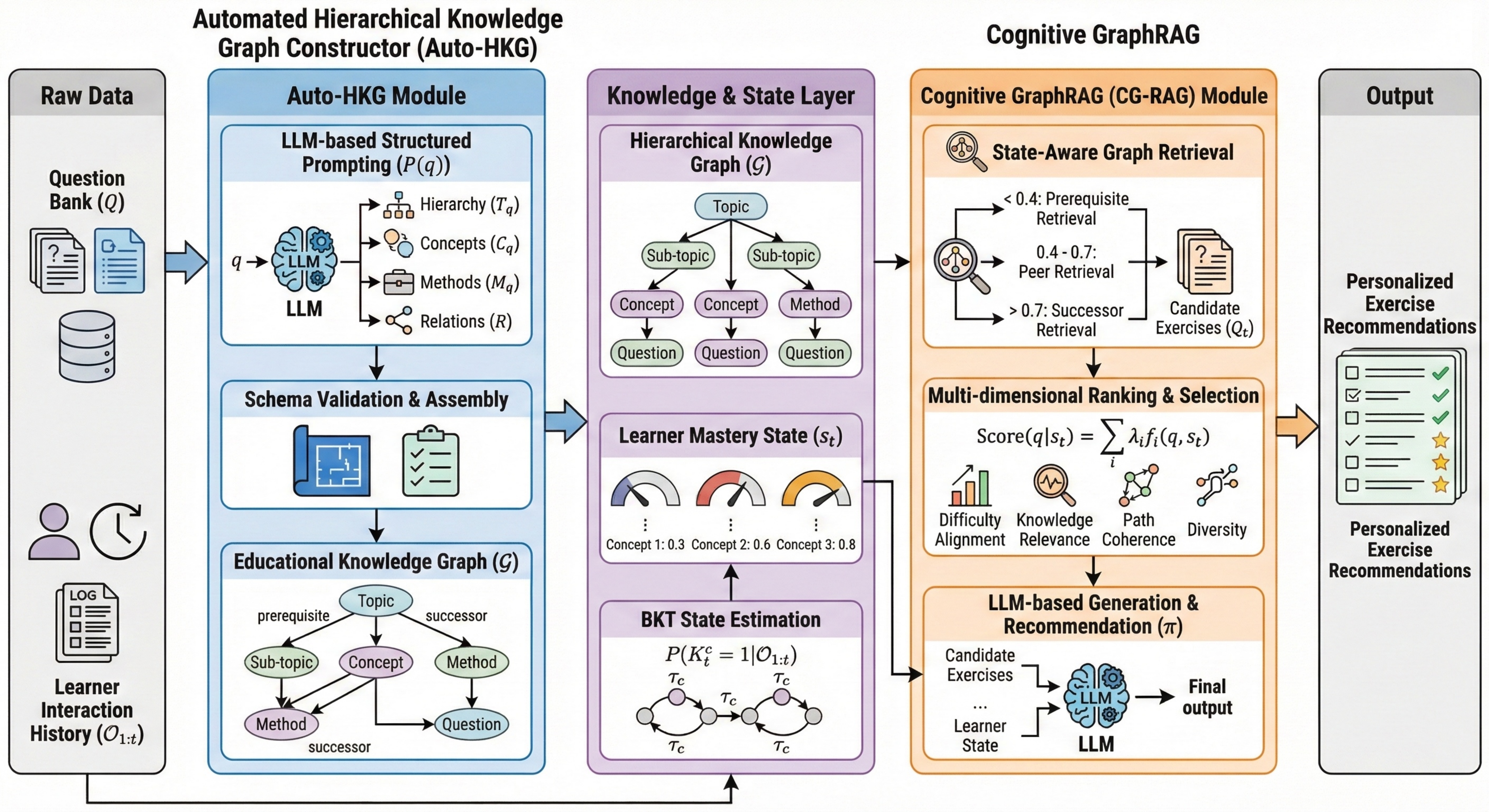}
    \caption{System architecture of the proposed framework, integrating automated hierarchical knowledge graph construction (Auto-HKG) with learner-aware graph reasoning and retrieval-augmented generation (CG-RAG) for personalized exercise recommendation.}
    \label{fig:pipeline}
\end{figure*}

\subsection{Problem Definition}

We focus on the personalized adaptive exercise recommendation and generation task in educational systems.
The goal is to recommend exercises that are pedagogically appropriate by jointly considering structured educational knowledge and the learner’s current mastery state.
Formally, let $Q$ denote the set of exercises and $C$ is the set of knowledge concepts.
We first construct an education knowledge graph $\mathcal{G}$ to encode semantic relations such as concepts hierarchies and prerequisite dependencies.
Then, the learner state is represented as a concept-level mastery vector $s_i\in [0, 1]$, where each entry reflects the learner's mastery of a knowledge concepts.
Given the knowledge graph $\mathcal{G}$, the learner state $S = [s_1, s_2, \cdots, s_{|C|}]$, and historical interactions, the system aims to learn a recommendation function
$
  \pi:(\mathcal{G}, S) \rightarrow Q,
$
which selects exercises whose associated concepts and prerequisite are aligned with the learner;s current mastery, enabling coherent and adaptive learning trajectories.

\subsection{Automated Hierarchical Knowledge Graph Constructor}

Constructing educational knowledge graphs remains challenging due to the hierarchical organization of educational content and complex prerequisite relationships among knowledge components.
Although knowledge concepts within a subject are usually well-defined, associating exercises with relevant concepts and modeling prerequisite relationships are non-trivial.
Existing approaches largely depend on expert-driven annotation or handcrafted rules, resulting in high maintenance costs and limited scalability as educational content evolves.
These limitations make it difficult for real-world systems to sustain high-quality knowledge graphs over time.

To address this problem, we propose the Automated Hierarchical Knowledge Graph Constructor (Auto-HKG), an LLM-based framework designed to automatically extract structured educational knowledge from question banks.
Let $Q = \{q_1, q_2, \cdots, q_n\}$ denotes the set of questions.
Auto-HKG employs structured prompting to guide the Large Language Model (LLM) toward producing graph-oriented outputs from the questions banks.
Specifically, given a question $q \in Q$, Auto-HKG uses a prompting function
$
  P(q) \rightarrow (T_q, C_q, M_q),
$
where $T_q$ denotes hierarchical categorizations (including coarse-grained and fine-grained topics), $C_q$ denotes the associated knowledge concepts, and $M_q$ denotes the problem-solving methods required by the question.
In addition, the model is instructed to infer prerequisite and successor relationships among concepts, yielding a set of semantic relations
$
  R \subseteq C \times C
$
which capture learning dependencies among knowledge components.
The extracted information is organized according to a predefined knowledge graph schema. Formally, Auto-HKG constructs an educational knowledge graph
$
  \mathcal{G} = (\mathcal{V}, \mathcal{E}),
$
where the node set $\mathcal{V} = Q \cup T \cup C \cup M$ includes questions, hierarchical categories, knowledge concepts, and problem-solving methods, and the edge set $\mathcal{E}$ encodes semantic relations such as prerequisite and successor links.
By imposing schema-level constraints on the LLM outputs, Auto-HKG ensures that the generated knowledge graph remains structurally coherent and amenable to downstream reasoning.
Auto-HKG is implemented as a pipeline that incrementally transforms raw educational resources into a structured hierarchical knowledge graph.
The pipeline integrates prompt-based extraction, schema validation, and graph assembly, allowing the system to continuously incorporate new questions and materials while maintaining graph consistency.
This modular pipeline design facilitates scalable knowledge graph construction and provides a foundation for learner-aware reasoning and personalized exercise generation in subsequent stages.

\subsection{Cognitive GraphRAG}

While Auto-HKG provides a structured and scalable representation of educational knowledge, effective personalization further requires integrating this static knowledge structure with the learner’s evolving cognitive state.
To this end, we introduce Cognitive GraphRAG (CG-RAG), a learner-aware reasoning and recommendation framework that performs state-driven graph reasoning and Retrieval-Augmented Generation (RAG) over the constructed knowledge graph.

CG-RAG first models the learner’s mastery state and its evolution over time. 
We denote the learner’s current cognitive state at time step $t$ as a concept-level mastery vector
$
  s_t = (s_t^1, s_t^2, \cdots, s_t^{|C|}) \in [0, 1]^{|C|},
$
where $s_t^c$ represents the learner’s mastery probability of knowledge concept $c\in C$ at time $t$.
Under a Bayesian Knowledge Tracing (BKT) formulation, for each concept $c$, the learner’s mastery is modeled as a latent binary variable $K^c_t \in \{0, 1\}$, indicating whether the concept is mastered.
The mastery probability corresponds to the $c$-th entry of the state vector:
$
  s_t^c = P(K^c_t = 1|\mathcal{O}_{1:t}),
$
where $\mathcal{O}_{1:t}$ denotes the observed learner–exercise interaction history up to time $t$.
Following the standard BKT state transition model, mastery evolution is governed by a concept-specific learning probability $\tau_c$.
Given an interaction outcome $o_t\in\{0, 1\}$ (incorrect or correct), the mastery state evolves as:
$
  P(K_{t+1}^c = 1| K^c_t = 0) = \tau_c,
$
$
  P(K_{t+1}^c = 1|K^c_t =1) =1.
$
Accordingly, the learner state is updated from $s_t$ to $s_{t+1}$ after each interaction.
Based on the estimated learner state $s_t$, CG-RAG adopts a mastery-driven graph retrieval strategy to guide subsequent reasoning and generation.
Specifically, when $s_t^c < 0.4$, the learner is considered to have insufficient mastery of the target concept.
In this case, CG-RAG prioritizes traversing prerequisite concepts of $c$ in the knowledge graph and retrieves exercises associated with these prerequisite nodes, aiming to reinforce foundational knowledge.
when $s_t^c > 0.7$, the learner is regarded as having achieved a relatively high level of mastery.
The retrieval strategy then shifts toward successor concepts of $c$, selecting exercises that support knowledge expansion and advanced learning.
For intermediate mastery levels, $0.4 \le s_t^c \le 0.7$, CG-RAG focuses on retrieving exercises associated with peer-level concepts or directly connected nodes in the graph, facilitating a smooth and stable learning progression.
This state-aware retrieval mechanism enables CG-RAG to adaptively navigate the knowledge graph according to the learner’s evolving mastery, ensuring that retrieved contexts and generated exercises are pedagogically aligned with the learner’s current learning needs.

Following state-aware graph retrieval, CG-RAG further performs fine-grained ranking over the retrieved candidate exercises using a multi-dimensional scoring function.
Let $\mathcal{Q}_t \subseteq \mathcal{Q}$ denote the candidate set obtained from graph traversal conditioned on the learner state $s_t$.
For each candidate exercise $q \in \mathcal{Q}_t$, the system computes an overall score
$
  \text{Score}(q|s_t) = \sum^{4}_{i=1}\lambda_if_i(q, s_t), 
$
where $f_{i}(\cdot)$ represents a scoring component and $\lambda_i$ denotes its corresponds weight.
The first component measures difficulty alignment between the exercise and the learner’s current mastery state, with weight $\lambda_1 = 0.4$.
This score is computed based on the distance between the estimated difficulty of exercise $q$ and the target difficulty interval induced by the learner’s mastery probability $s_t$, assigning higher scores to exercises whose difficulty better matches the learner’s current level.
The second component captures knowledge relevance, weighted by $\lambda_2 = 0.3$.
It is derived from the shortest path distance between the exercise-associated concepts and the target concept in the knowledge graph $\mathcal{G}$.
Shorter paths indicate stronger semantic relevance and yield higher scores.
The third component evaluates learning path coherence, with weight $\lambda_3 = 0.2$.
This term favors exercises involving several related concepts avoiding questions that are either overly narrow or excessively complex.
Finally, a diversity factor is incorporated with weight $\lambda_4 = 0.1$.
This component introduces a small amount of stochasticity into the ranking process, reducing redundancy among recommended exercises and preventing overly repetitive selections.
After scoring, CG-RAG conditions a large language model on the computed scores and the learner state $s_t$ to produce the final recommended exercise set from the candidate pool $\mathcal{Q}_t$.
By explicitly incorporating BKT-based mastery estimates and graph-aware relevance signals into the LLM’s input, the system generates exercise recommendations that align with the learner’s current proficiency while preserving logical coherence and progressive knowledge expansion along the knowledge graph.

\section{Experiments}

\subsection{Experiment Setup}

Following the problem setting and system design described in previous sections, we evaluate the proposed framework in the domain of high-school mathematics.
We collect a total of 1,007 exercises from standard high school mathematics textbooks, which serve as the input question bank for automated knowledge graph construction.
Based on the curriculum structure, we summarize 59 coarse-grained knowledge categories and 324 fine-grained knowledge concepts, corresponding to the hierarchical and conceptual nodes defined in the Auto-HKG schema.
We utilize the Pangu-7B~\cite{chen2025pangu} for structured knowledge extraction and retrieval-augmented generation.
%\cite{chen2025pangu,zeng2021pangu}

\begin{figure}[htbp]
    \centering
    \includegraphics[width=1.0\linewidth]{\picfolder 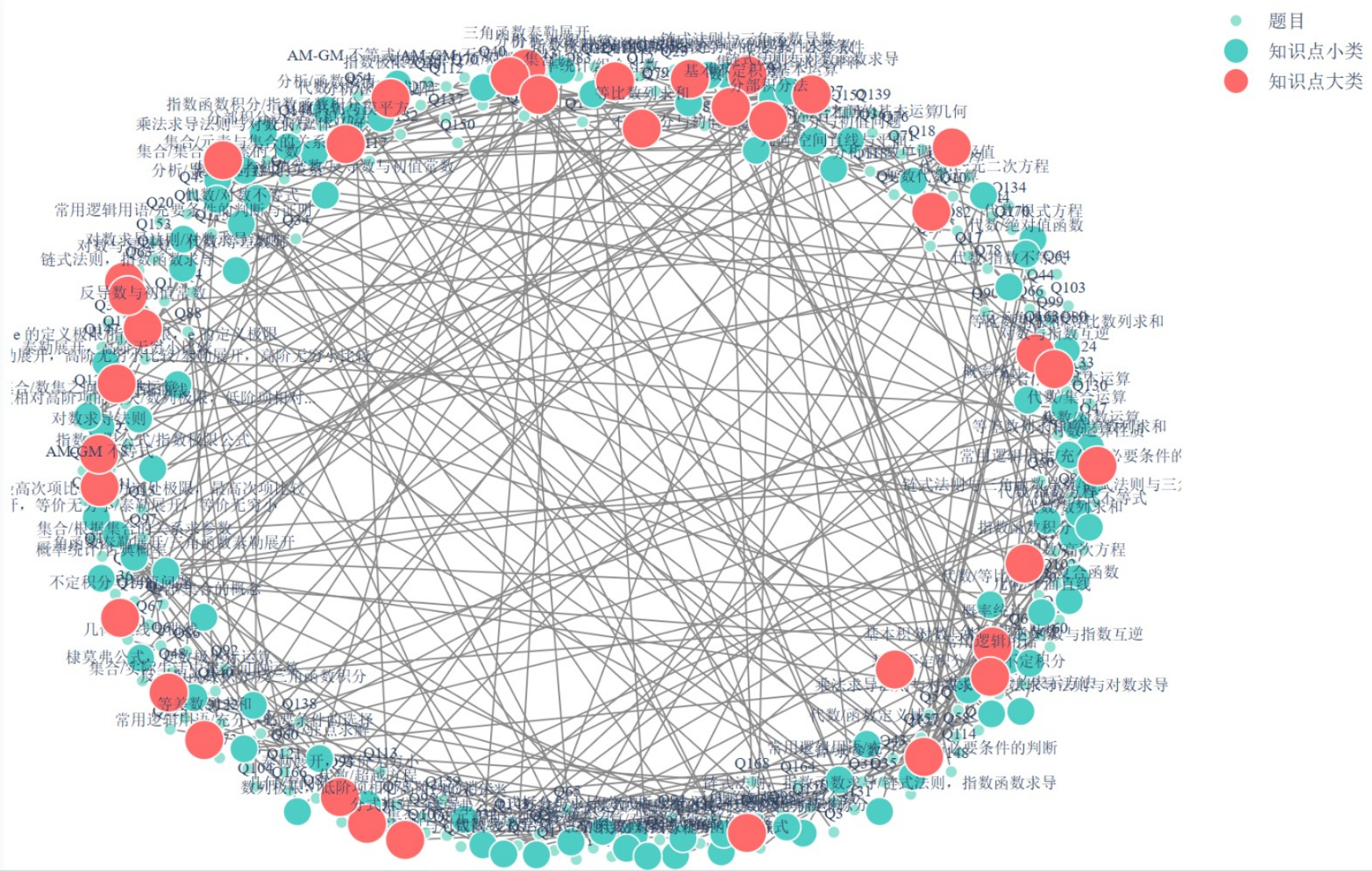}
    \caption{Visualization of the hierarchical educational knowledge graph automatically constructed by Auto-HKG.}
    \label{fig:vis_kg}
\end{figure}

\subsection{Knowledge Graph Construction Analysis}
For coarse-grained knowledge category extraction, Auto-HKG identifies a total of five major categories from the dataset.
% 需要说明具体数量
Upon manual verification against the standard high-school mathematics curriculum, three categories are found to be correctly defined.
The remaining category is summarized by the LLM as ``Analysis'', which does not correspond to an official coarse-grained category in high-school mathematics and should instead be categorized as ``Functions and Derivatives''.
Overall, Auto-HKG achieves an \textbf{75\%} accuracy in coarse-grained knowledge category definition for the high-school mathematics domain.
For fine-grained knowledge concept extraction, we manually verify the extracted concepts for 50 randomly sampled exercises.
Among them, 45 exercises are correctly assigned to appropriate fine-grained concepts, while 5 exercises are wrong (e.g., “sequence general term,” “basic inequality,” “inequality,” and “function limits”).
Overall, Auto-HKG achieves a fine-grained concept extraction accuracy of \textbf{90\%}, indicating that the effectiveness of LLM-based structured prompting in capturing detailed educational semantics.
In terms of efficiency, Auto-HKG demonstrates practical graph construction performance.
Using the Pangu-7B~\cite{chen2025pangu}, the system constructs knowledge graph nodes at an average speed of 1–3 seconds per node, including concept extraction and relation inference.
This efficiency enables scalable graph construction and supports iterative updates as question banks evolve.
As shown in Fig~\ref{fig:vis_kg}, the constructed knowledge graph reflects structured connections among coarse-grained knowledge categories, fine-grained concepts, and exercises, illustrating the hierarchical and relational organization learned by Auto-HKG.

\begin{figure}[htbp]
    \centering
    \includegraphics[width=1.0\linewidth]{\picfolder 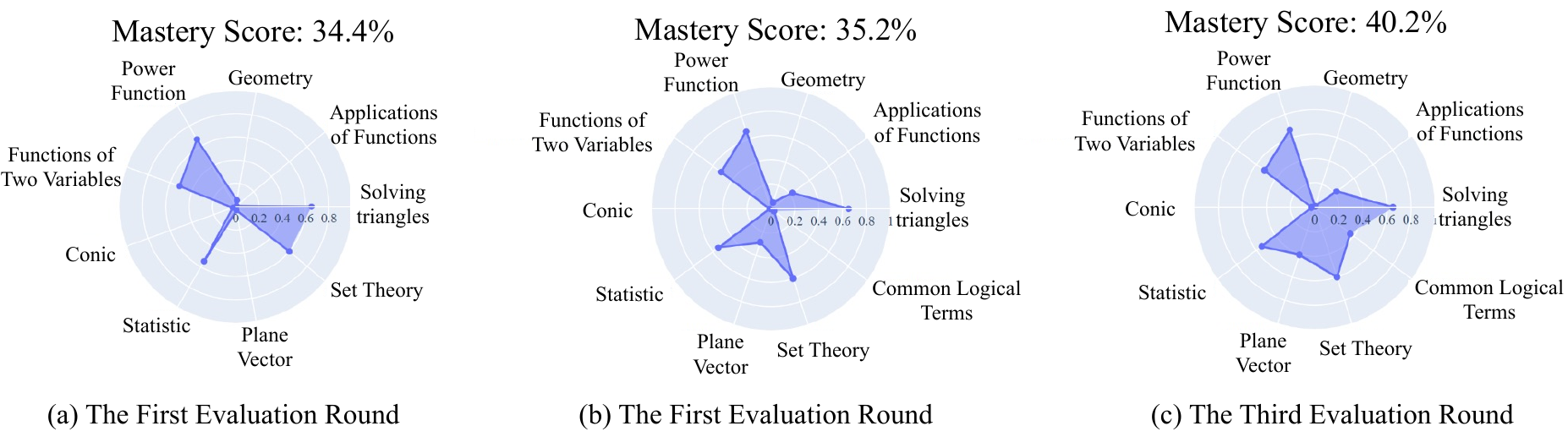}
    \caption{Visualization of the hierarchical educational knowledge graph automatically constructed by Auto-HKG.}
    \label{fig:rader}
\end{figure}

\subsection{Learner State–Aware Retrieval Behavior}
We analyze learner state evolution under the proposed system using consecutive evaluation rounds.
As shown in Fig.~\ref{fig:rader}, after system-guided exercise recommendation, the learner’s overall mastery score increases from 0.0 at initialization to 34.4\%, 35.2\%, and 40.2\% after the first, second, and third evaluation rounds, respectively.
Meanwhile, the number of mastered knowledge concepts increases from 0 to 11, and the cumulative number of attempted exercises grows from 0 to 30.
The corresponding knowledge radar charts indicate that mastery improvements are distributed across multiple knowledge concepts rather than concentrated on a single topic.
These observations suggest that learner state–aware retrieval enables consistent knowledge acquisition across evaluation rounds, supporting progressive learning aligned with the learner’s evolving mastery state.

\section{Conclusion}

In this paper, we presented Generative GraphRAG, a unified framework for automated knowledge modeling and learner-aware exercise generation in personalized education systems.
The framework integrates Auto-HKG for hierarchical educational knowledge graph construction and CG-RAG for mastery-driven graph reasoning and retrieval-augmented generation.
Through experiments in the high-school mathematics domain, we demonstrated that the proposed system construct educational knowledge graphs with reasonable accuracy and efficiency, and adapt exercise retrieval and generation according to learners’ evolving mastery states.
Our qualitative and quantitative analyses suggest that embedding learner state modeling into graph-aware retrieval provides a practical foundation for coherent and adaptive learning support.
Future work will explore broader domain coverage and larger-scale deployment.

\bibliographystyle{ACM-Reference-Format}
\bibliography{sample}

\end{document}